\documentstyle[12pt]{article}
\begin{document}
\input epsf
\begin{titlepage}
\begin{center}
\today     \hfill    WM-98-110\\

\vskip .25in

{\large \bf Decays of $\ell=1$ Baryons to $\Delta \gamma$}

\vskip 0.3in

Carl E. Carlson and Christopher D. Carone 

\vskip 0.1in

{\em Nuclear and Particle Theory Group \\
     Department of Physics \\
     College of William and Mary \\
     Williamsburg, VA 23187-8795}


        
\end{center}

\vskip .1in

\begin{abstract}
Recently we considered the electromagnetic decays of the orbitally-excited 
SU($6$) {\bf 70}-plet baryons in large-$N_c$ QCD, fitting to the helicity
amplitudes measured in photoproduction experiments.  Using the results
of this analysis, we predict the helicity amplitudes for the decays 
$N^* \rightarrow \Delta\gamma$ and $\Delta^* \rightarrow \Delta\gamma$. These 
decays can be studied at a number of new experimental facilities, and thus 
provide another nontrivial test of the lowest order large-$N_c$ predictions.
\end{abstract}
\end{titlepage}

\newpage
\renewcommand{\thepage}{\arabic{page}}
\setcounter{page}{1}
\section{Introduction} \label{sec:intro} \setcounter{equation}{0}

While the QCD gauge coupling is not perturbative at low energies, it is 
nonetheless possible to formulate an expansion in terms of the parameter 
$1/N_c$, where $N_c$ is the number of colors \cite{tHooft}.  In recent years,
effective field theories for baryons have been constructed that exploit this 
fact, allowing physical observables to be computed to any desired 
order in $1/N_c$.  For the ground state baryons, the SU(6) {\bf 56}-plet, 
the large-$N_c$ approach has been used successfully to study SU(6) 
spin-flavor symmetry \cite{DM,Jenk1,DJM1,CGO,Luty1}, 
masses \cite{DJM1,Jenk2,DJM2,JL}, magnetic 
moments \cite{DJM1,DJM2,JL,JM,Luty2,DDJM}, and axial current matrix 
elements \cite{DM,DJM1,DJM2,DDJM}.

Whether the large-$N_c$ framework works equally well in describing
the phenomenology of excited baryon multiplets is a question that is
under active investigation.  Recent attention has focused on the $\ell=1$ 
orbitally-excited baryons, the SU(6) {\bf 70}-plet for $N_c=3$.  
There have been studies of the masses \cite{Goity,CCGL}, strong 
decays \cite{CGKM,PY2}, axial current matrix elements \cite{PY1}, as well 
as the radiative decays of these states to ordinary nucleons \cite{CC}.  
However, the radiative decays to $\Delta$ final states have not been 
considered, and are of considerable interest, as we describe 
below.  These decays will be the main focus of this paper.

There is a good reason why much of the past literature \cite{old,gc} has 
focused on the radiative decays of excited baryons to nucleons rather than 
deltas: these are the decays for which there is experimental data.  The 
helicity amplitudes that describe the decays to nucleons are extracted 
experimentally by considering instead the time-reversed process, pion 
photoproduction. Information is available for the decays to nucleons simply 
because the fixed targets used in experiment are made of nucleons, not 
deltas.  To study the decays of excited baryons (in our case $N^*$'s or 
$\Delta^*$'s) to $\Delta \gamma$ not only requires that we produce enough 
excited states to make up for the small electromagnetic branching fractions, 
but also that we reconstruct a sufficient fraction of the events given the
final three decay products.  The 1998 Review of Particle 
Physics \cite{rpp98} lists no data for the $\Delta \gamma$ partial decay 
widths, nor for the more challenging helicity amplitudes, which require 
data on the angular distributions of the decays.

One reason why an analysis of the decays to $\Delta \gamma$ is of timely
interest is the possibility that the experimental situation may soon change, 
given, for example, the ongoing work at the Continuous Electron Beam 
Accelerator Facility (CEBAF).  CEBAF's high integrated beam luminosity, 
combined with the CEBAF Large Angle Spectrometer's (CLAS) efficiency for 
detecting photons in the forward direction makes study of the decays 
$N^*\rightarrow \Delta \gamma$  and $\Delta^* \rightarrow \Delta \gamma$ 
a possibility worthy of consideration. The Crystal Ball detector at 
Brookhaven may also allow study of the $\Delta \gamma$ decays, with 
excited baryons produced via a pion rather than a photon beam.

In this paper we will present our predictions for the $\Delta \gamma$ decay
amplitudes based on the large-$N_c$ operator analysis of Ref.~\cite{CC}.  The 
leading-order predictions following from single quark as well as single 
plus multiquark interactions are presented in algebraic and numerical 
form in the following section.  Study of the the $\Delta \gamma$ decays 
can give us further information on the significance of 
the multibody interactions that are included systematically in the 
large-$N_c$ approach, but are less frequently taken into account in quark 
model analyses.  While we will not embark on any detailed accelerator 
simulations to address the question of precisely how well the radiative 
decays to $\Delta$'s can be measured at various facilities, we will provide 
in the third section a discussion of how the amplitudes we predict can be 
extracted from the differential decay widths measured in experiment.  We 
hope this will provide strong motivation for experimenters to explore how 
well they can do. In the final section we summarize our conclusions.

\section{Decay Amplitudes} \label{sec:damps} \setcounter{equation}{0}

In this section, we present our predictions for the 
$N^*\rightarrow \Delta \gamma$ and $\Delta^* \rightarrow \Delta \gamma$ 
helicity amplitudes.  These follow directly from the large $N_c$ 
operator analysis of Ref.~\cite{CC}.   The formulation of an effective
theory for baryons in the large-$N_c$ limit has been discussed extensively 
in Refs.~\cite{CGO,CGKM}, so we will only provide a brief summary here:
Baryon states can be conveniently labeled by the SU(6)$\times$O(3) 
quantum numbers of their valence quarks.  For baryons of small total
spin within any given spin-flavor multiplet, this symmetry becomes
exact as $N_c \rightarrow \infty$, even if the valence quarks 
are light compared to $\Lambda_{QCD}$.  Thus, for the low spin states, this
spin-flavor space provides us with a basis for performing an operator 
analysis.  Operators with desired transformation properties may be formed 
by taking products of spin-flavor generators, O(3) generators, momenta, and 
polarizations of the states, and may involve one or more quark lines.
Operators that act on $n$ quark lines have coefficients suppressed by 
$1/N_c^{n-1}$, reflecting the $n-1$ gluon exchanges necessary to generate 
the operator in QCD. The $1/N_c$ power counting becomes nontrivial when one 
takes into account that compensating factors of $N_c$ may arise in the matrix 
elements of the operators, when a matrix element involves a coherent sum 
over O($N_c$) quark lines.  For low spin states, sums of the form 
\begin{equation}
\sum_\alpha \sigma^i_\alpha \,\,\, , 
\end{equation}
where $\sigma^i$ is a Pauli spin matrix, are incoherent, and of order one.  
On the other hand, sums of the form 
\begin{equation}
\sum_\alpha \lambda^a_\alpha \,\,\, \mbox{ or } \,\,\, 
\sum_\alpha \lambda^a_\alpha \sigma^i_\alpha \,\,\, ,
\end{equation}
where $\lambda$ is an SU(3) flavor matrix, are often coherent on at least 
some of the states.  To isolate the  corrections to a physical observable 
that appear at a given order in the $1/N_c$ expansion, one must take into 
account both the factors of $1/N_c$ that appear in the Lagrangian, as well 
as the compensating factors of $N_c$ that originate from taking matrix 
elements.

The analysis of radiative decays in Ref.~\cite{CC} focused on the one- 
and two-body operators that contribute to the helicity amplitudes at 
leading order, ${\cal O}$($N_c^0$).  In Coulomb gauge, the one-body
operators may be written:
\begin{equation}
a_1 Q_* \vec{\varepsilon}_m \cdot \vec{A} \,\,\, ,
\end{equation}
\begin{equation}
i b_1 Q_* \vec{\varepsilon}_m \cdot \vec{\nabla} 
(\vec{\sigma}_*\cdot\vec{\nabla}
\times \vec{A}) \,\,\, ,
\end{equation}
\begin{equation}
i b_2 Q_* \vec{\sigma}_*\cdot \vec{\nabla}(\vec{\varepsilon}_m 
\cdot \vec{\nabla}
\times \vec{A}) \,\,\, , 
\end{equation}
where the quark charge $Q$ is a matrix in SU($3$) flavor space, 
$Q=$diag$(2/3,-1/3,-1/3)$, $\vec{\varepsilon}_m$ is the polarization 
of the orbitally-excited quark, and an asterisk indicates that a given 
spin or flavor matrix acts only on the excited quark line.  We assume that the 
derivatives in these operators are suppressed by the scale $\Lambda_{QCD}$, 
which we have left implicit, for notational convenience.  These operators
are in one-to-one correspondence with the three operators included in
conventional quark model analyses \cite{close}; the precise relationship is 
given in Ref.~\cite{CC}. In addition to the operators above, a number of 
potentially coherent two-body operators were included in the analysis.  
The fits presented in Ref.~\cite{CC} demonstrated that the one-body 
operators provide a reasonable description of the experimental data, with 
only one of the two-body operators yielding a significant reduction in 
the $\chi^2$ of the fit.  Therefore, we include this two-body operator in 
the present analysis,
\begin{equation}
c_3 (\sum_{\alpha \neq *} Q_\alpha \vec{\sigma}_\alpha )\cdot \vec{\sigma}_* 
(\vec{\varepsilon}_m \cdot \vec{A}) \,\,\, ,
\end{equation}
where we have used the same labeling for undetermined coefficients as
in Ref.~\cite{CC}.  Note that this operator has exactly the type of 
spin-flavor sum that may be coherent on a large-$N_c$ baryon state.

The helicity amplitudes that we consider are defined by
\[
A_{-1/2} = K \, \xi \, \langle B^*, \,\, s_z=-\frac{1}{2} \,\, | H_{int}
| \,\, \gamma ,\,\, \epsilon_{+1}; \,\,B,\,\, s_z=-\frac{3}{2}\rangle
\]\[
A_{1/2} = K \, \xi \, \langle B^*, \,\, s_z=\frac{1}{2} \,\, | H_{int}
| \,\, \gamma ,\,\, \epsilon_{+1}; \,\,B,\,\, s_z=-\frac{1}{2}\rangle
\]\[
A_{3/2} = K \, \xi \, \langle B^*, \,\, s_z=\frac{3}{2} \,\, | H_{int}
|\,\, \gamma ,\,\, \epsilon_{+1}; \,\,B,\,\, s_z=\frac{1}{2}\rangle
\]\begin{equation}
A_{5/2} = K \, \xi \, \langle B^*, \,\, s_z=\frac{5}{2} \,\, | H_{int}
|\,\, \gamma ,\,\, \epsilon_{+1}; \,\,B,\,\, s_z=\frac{3}{2}\rangle \,\,\, ,
\label{eq:hampsdef}
\end{equation}
where the baryon states are relativistically normalized to $E/M_B$
particles per unit volume. Above, $s_z$ is the $z$-component spin in the 
$B^*$ rest frame, where $B$ represents either an $N$ or $\Delta$; $K$ is a 
kinematical factor given by $[4\pi\alpha m_{B^*}/(m^2_{B^*} - m^2_B)]^{1/2}$. 
The factor $\xi$ is the sign of the $\pi B B^*$ vertex that would appear in the
tree-level contribution to $\pi B \rightarrow B \gamma$; this renders our
sign conventions consistent with our previous work \cite{CC}.  Compared
to the $B^*\rightarrow N \gamma$ decays, two additional helicity amplitudes,
$A_{-1/2}$ and $A_{5/2}$, are required to compute physical observables.  The
dependence of various differential decay widths on the $A_\lambda$ are given in
the following section.

Compared to our previous study of the $B^*\rightarrow N\gamma$ decays, the 
computation involved in the current work is identical, except that (1)
we replace the spin-flavor wave function of the final state nucleon by that
of a $\Delta$, and (2) we compute one or two new matrix elements for each 
decay. The results we obtain for the $A_\lambda$ by evaluating the matrix 
elements of our four operators is presented in Table~\ref{table1}, for 
$N_c=3$.  There, $B_J^*$ and ${B_J^*}'$ represent baryon states with total 
spin $J$ and total quark spin $1/2$ and $3/2$, respectively. 

The physical baryon states, however, are not eigenstates of the total 
quark spin.  Two mixing angles are necessary to specify the $s=1/2$ and
$s=3/2$ nucleon mass eigenstates.  We define
\begin{equation}
\left[\begin{array}{c} N(1535) \\ N(1650) \end{array} \right] =
\left[\begin{array}{cc}  \cos\theta_{N1} & \sin\theta_{N1} \\
                       -\sin\theta_{N1} & \cos\theta_{N1} \end{array}\right] 
\left[\begin{array}{c} N^*_{1/2} \\ {N^*}'_{1/2}\end{array} \right]
\end{equation}
and
\begin{equation}
\left[\begin{array}{c} N(1520) \\ N(1700) \end{array} \right] =
\left[\begin{array}{cc}  \cos\theta_{N3} & \sin\theta_{N3} \\
                       -\sin\theta_{N3} & \cos\theta_{N3} \end{array}\right] 
\left[\begin{array}{c} N^*_{3/2} \\ {N^*}'_{3/2}\end{array} \right]  \,\,\, , 
\label{eq:tpt}
\end{equation}
as in Refs.~\cite{CGKM,CC}.  Using fit values for the operator coefficients
and mixing angles, we can make numerical predictions for the $A_\lambda$, for
each physical baryon state.

Our predictions are first presented in algebraic form in 
Table~\ref{table1}, in the limit of no mixing.  As in Ref.~\cite{CC}, we 
absorb any factors of momentum/$\Lambda_{QCD}$ that appear in the operators 
into our definitions of the fit coefficients; at leading order in $1/N_c$, 
these factors are multiplicative constants over the entire baryon multiplet.  
This redefinition is equivalent to replacing derivatives by unit vectors
$\hat{k}$. We present two sets of numerical predictions, corresponding 
to (i) a fit that includes only the one-body operators, and (ii) a fit that 
includes the one-body operators and the operator $c_3$. These are
shown in Tables~\ref{table2} and \ref{table3}, respectively.  While
$c_3$ and $a_1$ appear in the same combination for the 
$\Delta\gamma$ decays, the same is not true for the $N\gamma$ decays, and
thus the numerical fits are different. Note that 
the experimentally measured masses are used in evaluating the kinematical 
factor $K$.  Our predictions correspond approximately to the fits presented 
in Tables II and III of Ref.~\cite{CC}.  There, the one-body fit treated 
the mixing angles as free parameters, while the fit that included the operator 
$c_3$ held the mixing angles fixed at the values determined from a 
large-$N_c$ analysis of the decays $B^* \rightarrow B \pi$ \cite{CGKM}.  The 
mixing angles in both cases agreed within errors.  For the sake of
consistency, we present all our predictions with the mixing angles 
set to the values given in Ref.~\cite{CGKM}: $\theta_{N1}=0.61\pm 0.09$ and 
$\theta_{N3}=3.04\pm 0.15$.  

Our two sets of predictions are qualitatively similar.  There are
a number of cases where the central values of amplitudes are noticeably 
shifted by the inclusion of the operator $c_3$.  However, taking into
account the uncertainties in the fit parameters, these differences are 2-4 
standard deviation effects.  It is worth pointing out that new experimental 
data will lead to a reduction in the errors of our fit parameters, and hence 
to a clearer distinction between the predictions with and without the 
two-body operator effects.  Nonetheless, Table~\ref{table2} contains our 
most reliable predictions given the present data.  We now consider how
these amplitudes may be extracted from experiment.

\begin{table}[ht]
\begin{center}
\begin{tabular}{lcccc}
\\ \hline\hline
& $\Delta_{1/2}^{*+}$ & $\Delta_{1/2}^{*0}$ & $p^*_{1/2}$ & ${p^*}'_{1/2}$
\\ \hline
$\tilde{A}_{-1/2}$ 
& $\frac{2}{3\sqrt{3}}b_1$ & 0 & $-\frac{2}{3\sqrt{3}} b_1$ & 
$-\frac{1}{3\sqrt{3}}(3 a_1 - 2 b_1 +3 b_2 - 3 c_3)$ \\
$\tilde{A}_{1/2}$  & $-\frac{2}{9}(b_1+2 b_2)$ & 0 
& $\frac{2}{9}(b_1+2 b_2)$ & 
$-\frac{1}{9}(3 a_1-4 b_1 + b_2 - 3 c_3)$ \\ \hline
\end{tabular} 

\begin{tabular}{lcccc}
\\ \hline\hline
& $\Delta_{3/2}^{*+}$ & $\Delta_{3/2}^{*0}$ & $p^*_{3/2}$ & ${p^*}'_{3/2}$  
\\ \hline
$\tilde{A}_{-1/2}$ & $\frac{2\sqrt{2}}{3\sqrt{3}} b_1$
& 0 & $-\frac{2\sqrt{2}}{3\sqrt{3}}b_1$
& $\frac{2}{3\sqrt{15}}(3 a_1 + b_1 + 3 b_2 - 3 c_3)$ \\
$\tilde{A}_{1/2}$ & $\frac{2\sqrt{2}}{9}(b_1-b_2)$
& 0 & $-\frac{2\sqrt{2}}{9}(b_1-b_2)$ 
& $\frac{4}{9\sqrt{5}}(3 a_1 - b_1 + b_2 - 3 c_3)$ \\
$\tilde{A}_{3/2}$ & $-\frac{2\sqrt{2}}{3\sqrt{3}} b_2$
& 0 & $\frac{2\sqrt{2}}{3\sqrt{3}}b_2$ & $\frac{2}{3\sqrt{15}}
(3 a_1 - 3 b_1 -b_2 - 3 c_3)$ \\ \hline 
\end{tabular}

\begin{tabular}{lc} 
& ${p^*}'_{5/2}$ \\\hline\hline
$\tilde{A}_{-1/2}$ & $-\frac{1}{\sqrt{15}}(a_1 + 2 b_1 + b_2 - c_3)$ \\
$\tilde{A}_{1/2}$  & $- \frac{1}{3\sqrt{5}} (3 a_1 + 4 b_1 + b_2 - 3 c_3 )$ \\ 
$\tilde{A}_{3/2}$  
& $-\frac{\sqrt{2}}{3\sqrt{5}} (3 a_1 + 2 b_1 - b_2 - 3 c_3)$ \\
$\tilde{A}_{5/2}$  & $-\frac{\sqrt{2}}{\sqrt{3}}(a_1-b_2 - c_3)$ \\ \hline
\end{tabular}
\caption{Helicity amplitude predictions in terms of the operator
coefficients $a_1$, $b_1$, $b_2$ and $c_3$, in the case of no mixing.
Here $\tilde{A}_\lambda$ is defined by $A_\lambda = K\xi \tilde{A}_\lambda$.
Amplitudes related to these by isospin have not been displayed.}
\label{table1}
\end{center}
\end{table}

\begin{table}[ht]
\begin{tabular}{lcccc}
\\ \hline\hline
& $A_{-1/2}$ & $A_{1/2}$ & $A_{3/2}$ & $A_{5/2}$ \\ \hline
$\Delta^+(1620)$ & $-0.042\pm 0.005$ & $0.073\pm 0.006$ & - & - \\
$\Delta^0(1620)$ &    $0$   &  $0$    & - & - \\
$\Delta^+(1700)$ & $-0.054\pm 0.007$  & $0.000\pm 0.005$ & $0.055\pm 0.006$ \\
$\Delta^0(1700)$ &    $0$   &  $0$    & $0$ & - \\
$p(1535)$       & $0.108\pm 0.010$ & $0.004\pm 0.006$ & - & - \\
$p(1650)$       & $-0.063\pm 0.007$ & $-0.128\pm 0.007$ & - & - \\
$p(1520)$       & $0.062\pm 0.009$ & $-0.016\pm 0.006$ & $-0.090\pm 0.008$
& - \\
$p(1700)$       & $0.043\pm 0.008$ & $0.123\pm 0.007$ & $0.169\pm 0.008$ 
& - \\
$p(1675)$  & $0.024\pm 0.008$ & $-0.019\pm 0.009$ & $-0.113\pm 0.009$
& $-0.258\pm0.012$ \\ \hline
\end{tabular}
\caption{Helicity amplitude predictions, in GeV$^{-1/2}$, using parameter
values from a one-body operator fit, approximately that of Table~II in
Ref.~\protect\cite{CC} (see the text): $a_1=0.615\pm 0.028$, 
$b_1=-0.295\pm 0.038$, $b_2=-0.299\pm 0.032$, $\theta_{N1}=0.61$ (fixed), 
$\theta_{N3}=3.04$ (fixed).}
\label{table2}
\end{table}

\begin{table}[ht]
\begin{tabular}{lcccc}
\\ \hline\hline
& $A_{-1/2}$ & $A_{1/2}$ & $A_{3/2}$ & $A_{5/2}$ \\ \hline
$\Delta^+(1620)$ & $-0.042\pm 0.005$ & $0.074\pm 0.006$ & - & - \\
$\Delta^0(1620)$ &    $0$   &  $0$    & - & - \\
$\Delta^+(1700)$ & $-0.055\pm 0.007$  & $0.001\pm 0.005$ & $0.057\pm 0.006$ \\
$\Delta^0(1700)$ &    $0$   &  $0$    & $0$ & - \\
$p(1535)$       & $0.144\pm 0.013$ & $0.024\pm 0.008$ & - & - \\
$p(1650)$       & $-0.107\pm 0.012$ & $-0.156\pm 0.007$ & - & - \\
$p(1520)$       & $0.057\pm 0.009$ & $-0.024\pm 0.006$ & $-0.098\pm 0.008$
& - \\
$p(1700)$       & $0.089\pm 0.013$ & $0.177\pm 0.013$ & $0.218\pm 0.013$ 
& - \\
$p(1675)$  & $0.002\pm 0.009$ & $-0.060\pm 0.013$ & $-0.173\pm 0.015$
& $-0.337\pm0.020$ \\ \hline
\end{tabular}
\caption{Helicity amplitude predictions, in GeV$^{-1/2}$, using parameter
values from the four parameter fit given in Table~III of 
Ref.~\protect\cite{CC}: $a_1=0.816\pm 0.061$, $b_1=-0.299\pm 0.038$, 
$b_2=-0.308\pm 0.032$, $c_3=-0.072\pm 0.020$, $\theta_{N1}=0.61$ (fixed), 
$\theta_{N3}=3.04$ (fixed).}
\label{table3}
\end{table}

\section{Cross section formulas}

The total decay rate for $B^* \rightarrow \Delta + \gamma$ is

\begin{equation}
\Gamma_\gamma = {k^2 \over \pi} {2m_\Delta \over (2J+1) m_{B^*}}
      \sum_{\lambda=-1/2}^{\lambda=J} 
      \left| A_\lambda \right|^2 ,
\end{equation}
where $k$ is the momentum of the outgoing photon in the
excited baryon rest frame.  The formula is, of course, the
same as the analogous expression for $B^* \rightarrow N \gamma$, except 
for the substitution of $m_\Delta$ for $m_N$ in the numerator and the 
expanded range of $\lambda$.

The helicity amplitudes $A_\lambda$ can be found
separately with more detailed measurement.  We will record
some of the relevant formulas.  Our goal will be to show
explicitly that a set of measurements can lead to separation
of the $|A_\lambda|$, rather than to do an exhaustive analysis
of, for example, interference with the non-resonant background.

The excited baryon may be produced in a photonic or pionic
reaction,

\begin{eqnarray}
\gamma + N \rightarrow B_J^* \rightarrow \Delta + \gamma ,
\nonumber \\
\pi + N \rightarrow B_J^* \rightarrow \Delta + \gamma ,
\end{eqnarray}
and, in either case, the $B_J^*$ will be tensor polarized, at
least for $J \ne 1/2$.  For the case of the pionic reaction,
the pion brings in neither helicity nor angular momentum
projection along its direction of motion, so the helicity of
the excited baryon can only be $\pm 1/2$.  In the photonic or
Compton reaction, the initial state can have total helicity
$\pm 1/2$ and $\pm 3/2$, which is reflected in the
possibilities available to the excited  baryon.  The
probabilities of finding the differing helicities in the excited
baryon are, however, not the same but are given by 

\begin{eqnarray}
p_{1/2} &=& { | A_{1/2}(\gamma N \rightarrow B_J^*) |^2
            \over
          | A_{1/2}(\gamma N \rightarrow B_J^*) |^2  +
          | A_{3/2}(\gamma N \rightarrow B_J^*) |^2}  ,
\nonumber \\[1.2ex]
p_{3/2} &=& { | A_{3/2}(\gamma N \rightarrow B_J^*) |^2
            \over
          | A_{1/2}(\gamma N \rightarrow B_J^*) |^2  +
          | A_{3/2}(\gamma N \rightarrow B_J^*) |^2}  ,
\end{eqnarray}
for helicity magnitude $|\lambda_{B^*}| = 1/2$ and $3/2$,
respectively.  We will suppose that the helicity amplitudes
for
$\gamma N \rightarrow B_J^*$ are well measured, so that the
numbers $p_\lambda$ are known.

\begin{figure}  

\centerline{ \epsfxsize=4.5 in \epsfbox{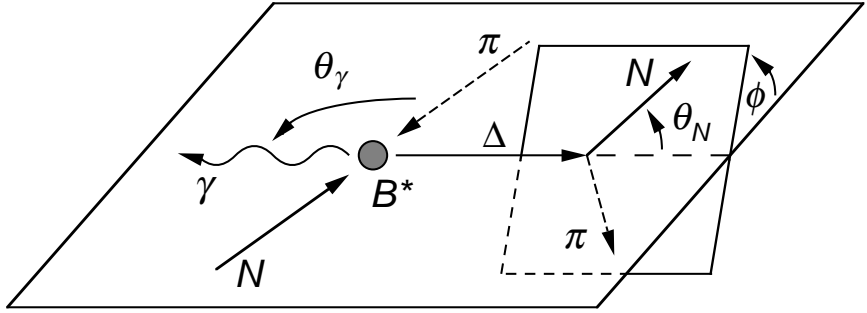}  }

\vglue .3 in

\caption{Production and decay of an excited baryon $B^*$ in
its own rest frame.   The angles $\theta_\gamma$,
$\phi$, and $\theta_N$ are indicated.  The $\theta_N$ used in
the text is $\theta_N$ boosted to the rest frame of the
$\Delta$.}

\label{2stepdecay}

\end{figure}

Because of the tensor polarization, the excited baryon does
not decay isotropically.  The angular distribution is given by
\begin{eqnarray}
{d\Gamma_\gamma \over d\Omega_\gamma} &&
                      (B_J^* \rightarrow \gamma \Delta) =
{k^2\over 4 \pi^2} {m_\Delta \over m_{B^*}}
                                       \nonumber \\ 
&&  \sum_{\lambda = -1/2}^{\lambda = J} 
    |A_\lambda|^2 
      \left\{ p_{1/2} \left[ |d^J_{1/2,\lambda}|^2  + 
                         |d^J_{-1/2,\lambda}|^2  \right]
        +
              p_{3/2} \left[ |d^J_{3/2,\lambda}|^2  + 
                         |d^J_{-3/2,\lambda}|^2  \right]
     \right\}       ,
\end{eqnarray}
Parity invariance has been used.  The 
$d^J_{M,\lambda} = d^J_{M,\lambda}(\theta_\gamma)$ are
elements of a matrix representation of rotations \cite{edmunds},
and $\theta_\gamma$ is the angle between the outgoing
photon and the incoming photon or pion in the rest frame of
the excited baryon (see Fig.~\ref{2stepdecay}).  Any
$A_\lambda$ not further specified is for $B^* \rightarrow
\gamma \Delta$.  If the excited baryon is produced  in the
pionic reaction, the above formula is valid with $p_{1/2}$
set to unity and $p_{3/2}$ set to zero.

Using the above expression,  one can separate $|A_{3/2}|$ and
$|A_{5/2}|$.   However, $|A_{-1/2}|^2$  and $|A_{1/2}|^2$ are
multiplied by the same kinematic factor, as one can see by
substituting $\lambda = \pm 1/2$ and using the symmetry of
the $d$-functions in the lower two indices.  Hence, some
further measurement is needed to separate them.

Recall that the $\Delta$ decays dominantly into $N\pi$, so
the full reaction is

\begin{eqnarray}
(\pi {\rm\ or\ } \gamma) + N \rightarrow B^* \rightarrow 
\gamma
 + \Delta 
\rightarrow \gamma + N + \pi    .
\end{eqnarray}

\noindent The angular distribution of the Delta decay
involves two more angles.  The decays of the~$B^*$ and
of the $\Delta$ each define a plane, and the angle between
them is the azimuthal angle $\phi$.  There is also a polar
angle
$\theta_N$, defined in the $\Delta$ rest
frame as the angle between the emerging $N$ and the $\Delta$
helicity axis inherited from the $B^*$ rest frame.  The
angular distribution of the
$\Delta$ decay depends on its helicity.  For example, if the
Delta has a definite helicity
$\lambda_\Delta$, the decay distribution is
proportional to

\begin{eqnarray}
{1\over 2} \left(1 + s_\lambda P_2 (\cos \theta_N) \right)
\end{eqnarray}

\noindent where $P_\ell$ is a Legendre polynomial, and

\begin{eqnarray}
s_\lambda = (-1)^{|\lambda_\Delta|-1/2}
       = \left\{ 
    \begin{array}{ll}
    +1, & \lambda = 1/2,3/2 \\
    -1, & \lambda = -1/2,5/2
    \end{array}
         \right.     .
\end{eqnarray}

\noindent The last part follows using 
$\lambda = 1 - \lambda_\Delta$.  All four helicity
amplitudes can be separated if one  measures the angular
distributions of the outgoing $N$ (or outgoing $\pi$) in
addition to that of the outgoing $\gamma$.

We will give the formulas for the double differential
cross sections, including an explicit evaluation of the
$d$-functions,  separately for $J = 1/2$,
$3/2$, and $5/2$:
\begin{eqnarray}
{d\Gamma_\gamma \left(B^*_{1/2}\right) \over 
               d \cos\theta_\gamma\, d\cos\theta_N } =
     {k^2 \over 4 \pi} {m_\Delta \over m_{B^*}}
     \left\{
        \left| A_{-1/2} \right|^2  
           \left( 1 - P_2 (\cos\theta_N) \right)
  +     \left| A_{ 1/2} \right|^2  
           \left( 1 + P_2 (\cos\theta_N) \right)
                                                  \right\}  ,
\nonumber
\\
\end{eqnarray}
\begin{eqnarray}
{d\Gamma_\gamma \left(B^*_{3/2}\right) \over 
               d\cos\theta_\gamma\, d\cos\theta_N } =
     {k^2 \over 4 \pi} {m_\Delta \over 2 m_{B^*}}
     &\Bigg\{&
          \left| A_{-1/2} \right|^2    
           \left( 1 + 
           [1 - 2p_{3/2}] P_2 (\cos\theta_\gamma) \right)
           \left( 1 - P_2 (\cos\theta_N) \right)
\nonumber \\
  &+&     \left| A_{ 1/2} \right|^2 
           \left( 1 + 
           [1 - 2p_{3/2}] P_2 (\cos\theta_\gamma) \right)
           \left( 1 + P_2 (\cos\theta_N) \right)
\nonumber \\
  &+&     \left| A_{ 3/2} \right|^2
           \left( 1 - 
           [1 - 2p_{3/2}] P_2 (\cos\theta_\gamma) \right)
           \left( 1 + P_2 (\cos\theta_N) \right)  \Bigg\},
\nonumber \\
  & &    
\end{eqnarray}
\noindent and
\begin{eqnarray}
&& {d\Gamma_\gamma \left(B^*_{5/2}\right) \over 
               d\cos\theta_\gamma\, d\cos\theta_N } =
      {k^2 \over 4 \pi} {m_\Delta \over 3 m_{B^*}} \times
\nonumber \\
&&\quad    \Bigg\{
        \left| A_{-1/2} \right|^2    
           \left( 1 + {8\over 7} 
           [1-{3\over 4}p_{3/2}] P_2 (\cos\theta_\gamma) 
      + {6\over 7} 
           [1-{5\over 2}p_{3/2}] P_4 (\cos\theta_\gamma)
                                                      \right)
           \left( 1 - P_2 (\cos\theta_N) \right)
\nonumber \\
&&\quad  +     \left| A_{ 1/2} \right|^2 
           \left( 1 + {8\over 7} 
           [1-{3\over 4}p_{3/2}] P_2 (\cos\theta_\gamma) 
      + {6\over 7} 
           [1-{5\over 2}p_{3/2}] P_4 (\cos\theta_\gamma)
                                                      \right)
           \left( 1 + P_2 (\cos\theta_N) \right)
\nonumber \\
&&\quad  +      \left| A_{ 3/2} \right|^2 
           \left( 1 + {2\over 7} 
           [1-{3\over 4}p_{3/2}] P_2 (\cos\theta_\gamma) 
      - {9\over 7} 
      [1-{5\over 2}p_{3/2}] P_4 (\cos\theta_\gamma) \right)
           \left( 1 + P_2 (\cos\theta_N) \right)
\nonumber \\
& &\quad +      \left| A_{ 5/2} \right|^2 
           \left( 1 - {10\over 7} 
           [1-{3\over 4}p_{3/2}] P_2 (\cos\theta_\gamma) 
      + {3\over 7} 
      [1-{5\over 2}p_{3/2}] P_4 (\cos\theta_\gamma) \right)
           \left( 1 - P_2 (\cos\theta_N) \right) \Bigg\} .
\nonumber \\
& &
\end{eqnarray}
\noindent Note that we have integrated over the azimuthal angle $\phi$ 
in deriving these decay widths.  We see that we can now extract the magnitude 
of each of the helicity amplitudes separately, the goal stated at the 
beginning of this section.  It is worth pointing out that we gain some,
but not complete, information on the relative signs of the amplitudes by
including the azimuthal angle dependence.  To determine the remaining signs 
requires a more detailed analysis, including for example polarizations 
and/or interference with the nonresonant background; we will consider these 
issues elsewhere. Again, to use these formulas for the pion induced 
reaction, set $p_{3/2}=0$.

\section{Discussion}

The advent of new experimental facilities makes the
measurement of excited baryon decay into $\Delta(1232) +
\gamma$ a real possibility.  In this paper, we have
considered decays of the {\bf 70}-plet.  There are 24 new
measurable amplitudes for {\bf 70} $\rightarrow \Delta
\gamma$, not counting amplitudes that are related using isospin 
invariance. 

Until now, the corresponding data on decays into $N \gamma$
has been obtained using time reversal invariance and
photoproduction.  This possibility doesn't exist for 
$\Delta \gamma$ decays, and the lack of data appears to have
engendered a paucity of theoretical study.  However,
measurements of the {\bf 70} $\rightarrow \Delta \gamma$ are
interesting for several specific reasons.

First, they are a test of the SU(6) symmetry that 
arises in the large-$N_c$ limit for baryons of low spin within
any given multiplet.  The $\Delta$ and the nucleon are both members
of the {\bf 56}-plet, and thus the {\bf 70} $\rightarrow \Delta \gamma$ 
amplitudes are predictable in terms of the same SU(6)-breaking parameters
that determine the {\bf 70} $\rightarrow N \gamma$ decays.   Considering
the one body operators, there are three such parameters, and the fit 
to the 19 $N \gamma$ decays is decently good.  If one assumes, as in 
the quark model, that only one-body decay operators are relevant,
then the  predictions for the $\Delta \gamma$ decays given in this paper  
provide 24 new opportunities to verify or vilify SU(6).  Our predictions
also provide a means of discerning effects of the most important two-body 
decay operators, those involving coherent sums over the large-$N_c$
baryon states. 

Secondly, the {\bf 70} $\rightarrow \Delta \gamma$ decays allow us to
test the assumption that two-body operators proportional to
quark spin sums have matrix elements that are incoherent.
Since the $\Delta$ spin is larger than that of the nucleon, it is
possible that these subleading effects are not sufficiently suppressed
for $N_c=3$ to justify the large-$N_c$ operator analysis. Then we might
encounter large corrections not present in the $N\gamma$ decays. Such
large multibody operator effects would lead to significant deviations
from the predictions presented here, as well as a noticeable
breakdown of the naive quark model.

\begin{center}              
{\bf Acknowledgments} 
\end{center}
We thank David Armstrong, Nathan Isgur, and Ron Workman for useful 
comments. CDC thanks the National Science Foundation for support under 
grant PHY-9800741. CEC thanks the NSF for support under grant PHY-9600415.


\begin{thebibliography}{99} 
\frenchspacing

\bibitem{tHooft} G. 't Hooft, Nucl.\ Phys.\ {\bf B72}, 461 (1974).

\bibitem{DM} R. Dashen and A. V. Manohar, Phys.\ Lett.\ {\bf B315},
425 (1993); {\it ibid.}, 438 (1993).

\bibitem{Jenk1} E. Jenkins, Phys.\ Lett.\ {\bf B315}, 431 (1993).

\bibitem{DJM1} R. Dashen, E. Jenkins, and A. V. Manohar, Phys.\ Rev.\
D {\bf 49}, 4713 (1994).

\bibitem{CGO} C. D. Carone, H. Georgi, and S. Osofsky, Phys.\ Lett.\
{\bf B322}, 227 (1994).

\bibitem{Luty1} M. A. Luty and J. March-Russell, Nucl.\ Phys.\ {\bf
B426}, 71 (1994).

\bibitem{Jenk2} E. Jenkins, Phys.\ Lett.\ {\bf B315}, 441 (1993).

\bibitem{DJM2} R. F. Dashen, E. Jenkins, and A. V. Manohar, Phys.\
Rev.\ D {\bf 51}, 3697 (1995).

\bibitem{JL} E. Jenkins and R. F. Lebed, Phys.\ Rev.\ D {\bf 52}, 282
(1995).

\bibitem{JM} E. Jenkins and A. V. Manohar, Phys.\ Lett.\ {\bf B335},
452 (1994);

\bibitem{Luty2} M. A. Luty, J. March-Russell, M. White, Phys.\ Rev.\ D
{\bf 51}, 2332 (1995).

\bibitem{DDJM} J. Dai, R. Dashen, E. Jenkins, and A. V. Manohar,
Phys.\ Rev.\ D {\bf 53}, 273 (1996).

\bibitem{Goity} J. L. Goity, Phys.\ Lett.\ {\bf B414}, 140 (1997).

\bibitem{CCGL} C. E. Carlson, C. D. Carone, J. L. Goity, and R. F. Lebed,
hep-ph/9807334, submitted to Phys. Lett. B.

\bibitem{CGKM} C. D. Carone, H. Georgi, L. Kaplan, and D. Morin,
Phys.\ Rev.\ D {\bf 50}, 5793 (1994).

\bibitem{PY2} D. Pirjol and T.-M. Yan, Phys.\ Rev.\ D {\bf 57}, 5434
(1998).

\bibitem{PY1} D. Pirjol and T.-M. Yan, Phys.\ Rev.\ D {\bf 57}, 1449
(1998).

\bibitem{CC} C. E. Carlson and C. D. Carone, Phys.\ Rev.\ D {\bf 58}
053005 (1998).

\bibitem{old} See, for example, L. A. Copley, G. Karl and E. Obryk, 
Phys.\ Lett.\ {\bf B29}, 117 (1969); Nucl.\ Phys.\ {\bf
B13}, 303 (1969).

\bibitem{gc} For an exception, see F. J. Gilman and I. Karliner,
 Phys.\ Rev.\ D {\bf 10}, 2194 (1974).

\bibitem{rpp98} Particle Data Group, C. Caso, {\em et al.}, 
Eur. Phys. J. {\bf C3} 1, (1998).

\bibitem{close}
F.E. Close, {\em Quarks and Partons}, Academic Press, London, 1979.

\bibitem{edmunds}
A. R. Edmonds, {\em Angular Momentum in Quantum Mechanics}, Princeton
University Press, Princeton, 1960.

\end{thebibliography}
\end{document}